\documentclass[%
 reprint,
 amsmath,amssymb,
 aip,
 rsi,
]{revtex4-2}

\usepackage{graphicx}
\usepackage{dcolumn}
\usepackage{bm}
\usepackage{hyperref}
\usepackage{xcolor}
\usepackage[normalem]{ulem}

\hypersetup{colorlinks=true,linkcolor=cyan,citecolor=cyan,filecolor=cyan,urlcolor=cyan}

\begin{document}

\preprint{APS/123-QED}

\title{Two-dimensional plasma density evolution local to the inversion layer during sawtooth crash events using Beam Emission Spectroscopy }

\author{Sayak Bose}
\email{sbose@princeton.edu}
\affiliation{Princeton Plasma Physics Laboratory, Princeton, New Jersey 08540, USA}

\author{William Fox}%
\email{wfox@pppl.gov}
\affiliation{Princeton Plasma Physics Laboratory, Princeton, New Jersey 08540, USA}

\author{Dingyun Liu}%
\affiliation{Princeton University, New Jersey 08540, USA}

\author{Zheng Yan}
\affiliation{University of Wisconsin-Madison, Madison, Wisconsin, USA}%

\author{George McKee}
\affiliation{University of Wisconsin-Madison, Madison, Wisconsin, USA}%

\author{Aaron Goodman}
\affiliation{Princeton University, New Jersey 08540, USA}

\author{Hantao Ji}
\affiliation{Princeton Plasma Physics Laboratory, Princeton, New Jersey 08540, USA}
\affiliation{Princeton University, Princeton, New Jersey 08540, USA}


\date{\today}

\begin{abstract}

We present methods for analyzing Beam Emission Spectroscopy (BES) data to obtain the plasma density evolution associated with rapid sawtooth crash  events at the DIII-D tokamak. BES allows coverage over a 2-D spatial plane, inherently local measurements, with fast time responses, and therefore provides a valuable new channel for data during sawtooth events.  A method is developed to remove sawtooth-induced edge-light pulses contained in the BES data. The edge light pulses appear to be from the $\rm{D}_{\alpha}$ emission produced by edge recycling during sawtooth events, and are large enough that traditional spectroscopic filtering and data analysis techniques are insufficient to deduce physically meaningful quantities.  A cross-calibration of 64 BES channels is performed using a novel method to ensure  accurate measurements.  For the large-amplitude density oscillations observed, we discuss and use the non-linear relationship between BES signal $\delta I/I_{0}$ and plasma density variation $\delta n_{e}/n_{e0}$. 2-D BES images cover a 8~cm~$\times$~20~cm region around the sawtooth inversion layer and show large-amplitude 
density oscillations, with additional significant spatial variations across the inversion
layer, which grows and peaks near the time of the temperature crash.
The edge light removal technique and method of converting large-amplitude $\delta I/I_{0}$ to $\delta n_{e}/n_{e0}$ presented here may help analyze other impulsive MHD phenomena in tokamaks.
\end{abstract}

\maketitle


\section{\label{sec:Introduction}Introduction}

Sawtooth oscillations\cite{Goeler1974} are internal relaxation events in a tokamak that lead to a rapid drop of core electron temperature. A significant question for sawtooth oscillations is what is the cause of the short crash time. In the traditional Kadomtsev model\cite{Kadomtsev1975}, the crash time is related to how fast reconnection can occur to re-arrange the magnetic field. Several possible competing mechanisms have been proposed for the fast crash, including two-fluid effects at the reconnection layer\cite{AydemirPoF1992,WangPRL1993,KlevaPoP1995,FoxPRL2017}, plasmoid instability\cite{loureiro2007instability,bhattacharjee2009fast,gunter2014fast}, and interchange instability\cite{wesson1986sawtooth,jardin2020new}. 

The understanding of the sawtooth crash phenomena has  improved with the development of new measurement capabilities.
Current profile measurements\cite{levinton1993q}, reconstruction of 2-D temperature profile using soft X-ray (SXR)\cite{nagayama1988soft} and electron cyclotron emission (ECE)\cite{nagayama1996tomography} tomography, direct imaging of 2-D electron temperature using electron cyclotron imaging \cite{park2019newly} (ECEI) have provided valuable information to differentiate between various models. For example, measurements of 2-D electron temperature using ECEI in TEXTOR\cite{park2006comparison,park2006observation} supported the occurrence of magnetic reconnection, while the SXR tomography diagnostic in JET\cite{granetz1988x} showed the presence of the interchange mode.

Beam emission spectroscopy\cite{FonckRSI1990} (BES) is an active plasma diagnostic that can  measure the time evolution of the plasma density in a 2-D plane. A high-energy hydrogenic neutral beam is injected into the plasma, and the associated emission from the collisionally-excited  neutral beam fluorescence is observed. The intensity of light emission is related to the plasma density via atomic physics. The valuable insights obtained from the localized density information supplied by the BES in various experiments\cite{mckee2006high, mckee2003turbulence, schlossberg2006velocity,yan2011high,yan2014observation} motivate further development of BES setups and analysis techniques to measure the plasma density during sawtooth events. The plasma density is a relatively unexplored measurement channel during sawtooth events and may show complementary physics to the 
temperature channel, which is dominated by the sawtooth-driven 
heat transport.  Observing the plasma density may therefore help determine which of several processes drive the sawtooth \cite{AydemirPoF1992,WangPRL1993,KlevaPoP1995,FoxPRL2017,loureiro2007instability,bhattacharjee2009fast,gunter2014fast,wesson1986sawtooth,jardin2020new}.

In this paper, we develop techniques for direct observation of plasma density local to the q=1 layer during sawtooth events using BES.   The most significant challenge we overcome is to isolate the core BES signals from stray “edge” light.  In typical DIII-D experiments, the spectral Doppler shift induced by a beam of velocity, $v/c \sim 0.01$ ($\sim$ 80 keV deuterium atoms), is a few nm, which is sufficient under usual applications to  spectroscopically isolate the beam emission from the $\rm{D}_{\alpha}$ emission produced by edge recycling, enabling accurate measurement of the core density fluctuations\cite{Mckee1999}. 
However, during the sawtooth events studied here, the $\rm{D}_{\alpha}$ edge light is exceptionally high in magnitude.
The edge light is transmitted to the BES photodiodes despite the high attenuation through the optical interference filter. The edge light temporally overlaps with the neutral-beam-driven core
emission, complicating the interpretation.  This motivates the present paper, in which we describe a technique to isolate and remove the undesired edge light from BES data to isolate the core BES signals. 
The technique may also be valuable to analyze plasma density evolution during other impulsive MHD events such as edge localized modes\cite{leonard2014edge}. Other active spectroscopic diagnostics employing a neutral beam may also find our technique to be useful.   

The scope of this paper is to discuss, in detail, the BES analysis procedure and a sample of measurements of the 2-D in-plane plasma density during sawtooth events. A large-amplitude density oscillation, likely associated with rotating a $(m,n) = (1,1)$ mode, is observed to grow near the onset of the crash in core electron temperature $T_{e,\rm{core}}$. This mode reaches its maximum amplitude at the latter end of the crash, after which it decays over a few cycles. In addition, a density gradient in the $R$-$Z$ plane across the $q=1$ surface is found to be associated with a sawtooth crash. A detailed physics analysis and study of multiple events will be pursued in upcoming publications.

The rest of the paper is organized as follows: Section~\ref{sec:overview} gives a brief overview of sawtooth oscillations and how spatially-resolved density measurements can help to constrain the physics. 
The experimental  setup is described in  Section~\ref{sec:experimental_set_up}. The method for BES analysis for sawtooth events and analysis of experimental results is presented in Section~\ref{sec:result_analysis}.  A discussion and summary follow this in Section~\ref{sec:discussion}.

\section{\label{sec:overview}Overview}

Sawtooth oscillations \cite{hastie1997sawtooth,chapman2010controlling} are a periodic relaxation of  $T_{e,\rm{core}}$ in tokamaks. These oscillations are characterized by a slow build-up of $T_{e,\rm{core}}$ followed by a rapid ``crash" phase, so that the
temporal evolution over several cycles appears sawtooth-like.  According to most models, the sequence of events leading to a crash are as follows: Typically, tokamaks have a peaked temperature profile. This supports a peaked toroidal current because of a higher conductivity at the center of the plasma. A higher toroidal current further heats the core increasing the $T_{e,\rm{core}}$ in a positive feedback loop. Enhancement of the toroidal current increases the poloidal magnetic field lowering the $q $ value near the core, where $q = \left \langle rB_\phi / R B_\theta \right \rangle$ is the inverse rotational transform of the magnetic field, and angle brackets refer to the flux surface average\cite{wesson2011tokamaks}. When $q$ in the core becomes less than unity,  current\cite{Kadomtsev1975} or pressure\cite{wesson1986sawtooth} driven MHD instabilities can grow.  These instabilities relax the temperature and current profiles, causing the cycle to begin anew.

The open physics question is, how a MHD mode or modes can cause a rapid crash in $T_{e,\rm{core}}$. According to models based on magnetic reconnection, a growing kink mode causes magnetic reconnection at $q = 1$ surface, which rearranges the magnetic field and relaxes the  temperature via fast transport along the newly 
reconnected field lines. However, as calculated by the Sweet-Parker model, resistive reconnection is not fast enough to explain the observed short crash time.  Extensions to reconnection theory have been proposed
to overcome this shortcoming. Two-fluid effects about a single reconnection current sheet or development of multiple plasmoid structures in the reconnection layer due to secondary tearing of the current sheet can speed up magnetic reconnection and explain the fast crash time\cite{AydemirPoF1992,gunter2014fast}.   2-D density measurements at the $q=1$ sawtooth inversion region would enable the detection of density structures that may be relevant to confirming these reconnection models. A signature of the two-fluid effect would be a quadrupolar variation of density in the inversion region i.e. two regions about the current sheet near the X-point with a positive density perturbation, and the other two with a negative\cite{KlevaPoP1995,FoxPRL2017,Bose2020,2020APSDPPJ11004B}.   The plasma density evolution may alternatively provide evidence for plasmoid reconnection\cite{gunter2014fast}. An alternative explanation of a sawtooth event that does not depend on reconnection predicts the crash is caused by 
higher-mode-number pressure-driven interchange modes\cite{jardin2020new}.   In this case, a 2-D  density measurement in the $R$-$Z$ plane may detect such higher-order MHD modes. We note that depending on the line of sight, interferometer data may exhibit signs of density fluctuations\cite{chapman2010controlling} as well.

\section{\label{sec:experimental_set_up}Experimental Setup}

\begin{figure}[h]
\includegraphics[scale=0.65]{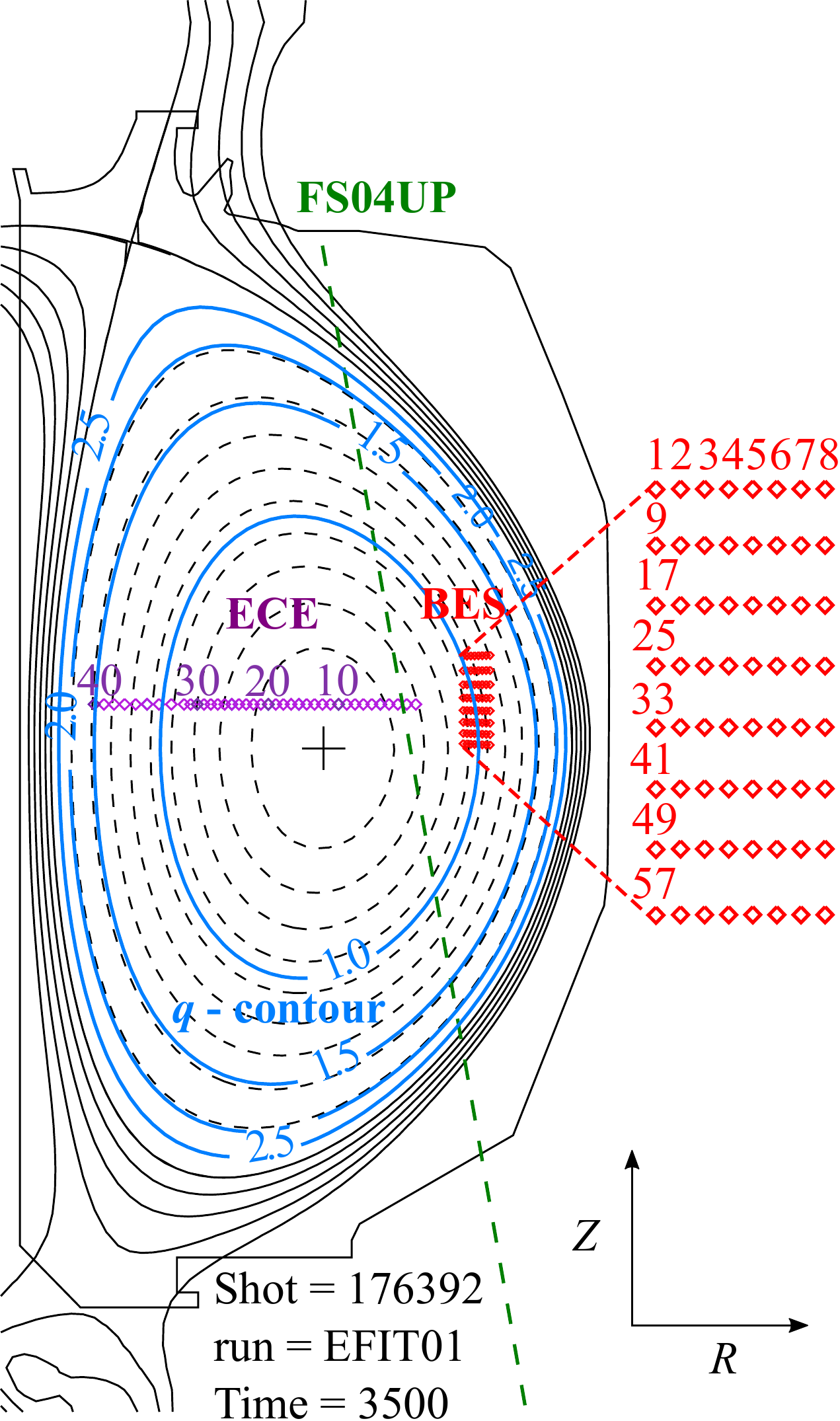}
\caption{\label{fig:setup} $R$-$Z$ plane of DIII-D showing the $q$-contours, BES channels, ECE channels and filterscope (FS04UP).}
\end{figure}

Experiments to measure the time resolved 2-D density evolution during sawtooth oscillations were conducted in DIII-D with a setup allowing for BES\cite{Mckee1999,mckee2010wide} measurements  at and around the $q=1$ surface. In order to shift the $q=1$ surface to a suitable location for the BES, a relatively low $q_a \sim 2.2$  DIII-D discharge was adopted in an L-mode plasma, where the toroidal magnetic field was  $\rm{B}_{\rm{T}}=$1.52~T, and the plasma current $I_{\rm{p}}$ was 1.76 MA.

The BES diagnostic uses a hydrogenic neutral beam of 45--80 keV energy (was 55 keV in our experiment) that is injected into the plasma by a neutral beam source. As neutrals in the beam collide with plasma electrons, ions, and impurities, a fraction of the beam neutrals enter the n = 3 state via direct excitation or cascade processes. Transitions from $\rm{n}=3$ to $\rm{n}=2$ state causes emission near $\lambda_0 = 656.1~\rm{nm}$. A rather high velocity of the neutral beam, $v/c \sim 0.01$, causes the emission manifold to be blue shifted to near $653 - 655~\rm{nm}$\cite{Mckee1999}. This blue shift  is sufficient to isolate the neutral beam fluorescence  from the thermal $\rm{D}_{\alpha}$ emission produced by edge recycling in most experiments using customized interference filters. The measured light intensity fluctuations are related to the plasma electron and ion density fluctuations through atomic physics of the beam excitation process and weakly dependent on other beam and plasma parameters\cite{FonckRSI1990,HutchinsonPPCF2002}.

At DIII-D, a 2-D BES system has been implemented which measures the beam fluorescence 
using 64 spatial channels\cite{Mckee1999, mckee2006high}.  A flexible fiber array mount  allows for rapid and easy reconfiguration of the 64 channels for the scientific needs of a given experiment. For this experiment, the array was organized in an $8\times8$ configuration for a total coverage of $8 \times 20$~cm in the radial-poloidal plane.  The location of the BES channels is shown in Figure \ref{fig:setup}, which shows an EFIT plasma equilibrium reconstruction along with several other diagnostics discussed in the present paper. The contour of the $q=1$ surface passes through the area sampled by BES.  
The BES measurement volume is located at a toroidal angle of $\sim140^{\circ}$. 
The BES sightlines are approximately tangent to flux surfaces and are angled to match the dominant magnetic field pitch angle to provide good spatial resolution perpendicular to the field lines. Each BES channel integrates emission over an area of approximately $1 \times 1.3\; \rm{cm}$ region in the $R$-$Z$ plane, though the detailed point-spread-function is calculated for each channel and shot based on position, equilibrium, and profiles using diagnostic geometry\cite{shafer2006spatial}.

The light acquired by the collection optics is converted to voltage signals by photodiodes \cite{Mckee1999,fonck1992low,gupta2004enhanced}. The high and low frequency components of the BES photodiodes signal are saved separately to provide extra bit resolution,
in what we refer to below as the ``fast'' and ``slow'' BES channels, respectively.  In particular, the fast channels are AC-coupled, and pass frequencies from $\sim1~\rm{kHz}$ up to a cutoff frequency of 425~kHz.

To complement the BES measurements, several
other diagnostics were used (Ref.~\cite{boivin2005diii} and references therein).  The electron temperature was measured using a 40 channel electron cyclotron emission radiometer located at a toroidal angle of $81^{\circ}$. The layout of the ECE channels is shown in Figure \ref{fig:setup}. We have used the channel 10 to measure the $T_{e,\rm{core}}$  which was found to be $\sim3~\rm{keV}$.

The electron density near the $q=1$ surface is measured using a multichannel Thomson scattering diagnostic. Since the time response of the Thomson scattering diagnostic is not fast enough to resolve density fluctuations owing to sawtooth oscillations, this diagnostic gives a local time-averaged electron density, $n_{e0}$. We compared the $n_{e0}$ measured using Thomson scattering at multiple locations in the neighborhood of the plasma sampled by BES. We did not observe any significant spatial variation of $n_{e0}$ and the magnitude of $n_{e0}$ is  $\sim 3.5 \times 10^{13}\; \rm{cm^{-3}}$. %

Magnetic field fluctuations, $dB/dt$, are measured using  B-dot probes which are located external to the plasma at various toroidal angles. Here we use $dB/dt$ measured by a high frequency B-dot probe located at a toroidal angle of $150^{\circ}$. 

A filterscope called FS04UP is used for line-integrated measurements of edge $\rm{D}_{\alpha}$  spectral line emission.  The line of sight of the FS04UP is shown by the green dashed line in Figure \ref{fig:setup}.

\section{\label{sec:result_analysis}Results and Analysis}
\subsection{Raw data and preprocessing}
\begin{figure}
    \centering
    \includegraphics[scale=0.88]{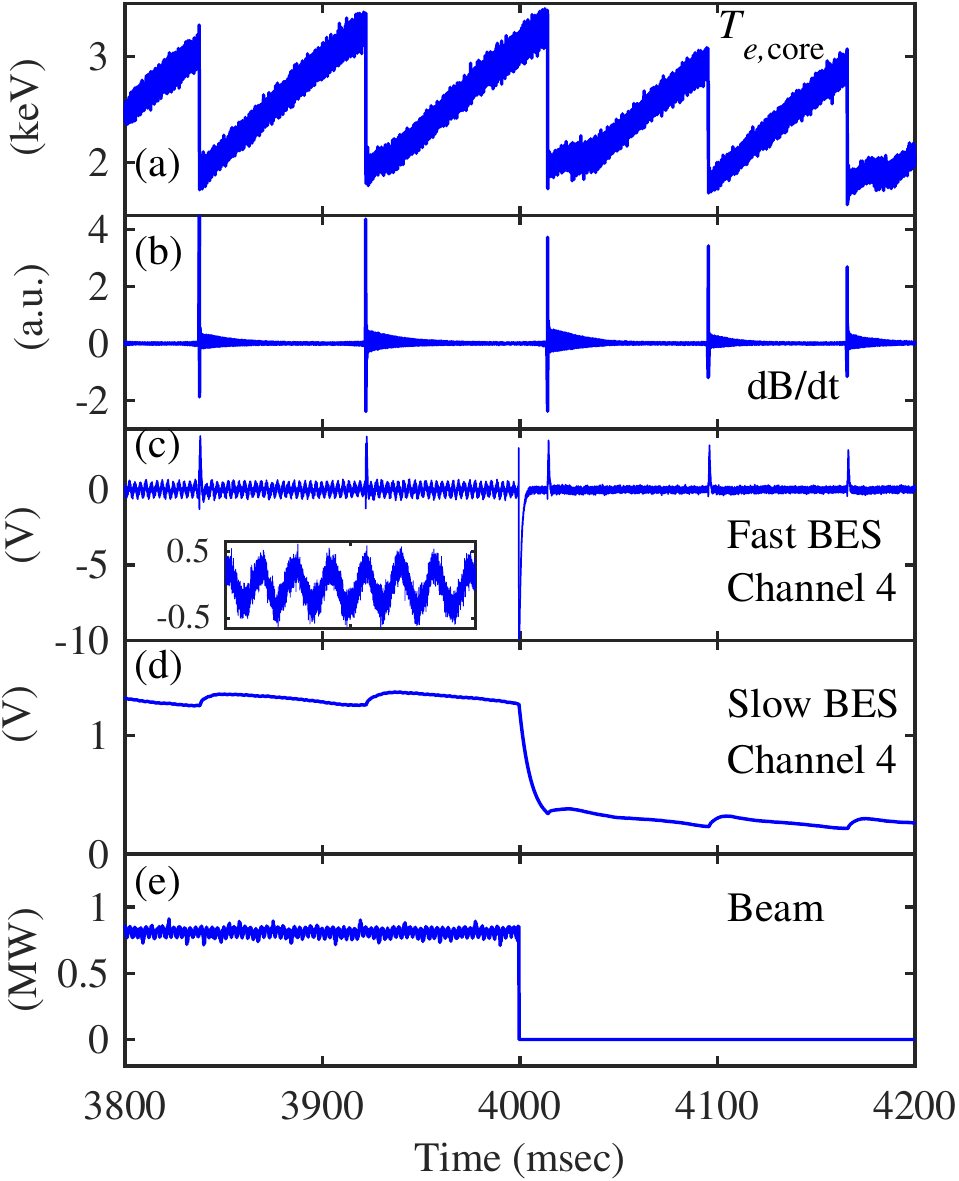}
    \caption{Time evolution of several quantities during
    a series of sawtooth events.  (a) $T_{e,\rm{core}}$ 
    measured by ECE, (b) B-dot probe showing a magnetic activity during temperature crash. Temporal variation of the (c) fast and (d) slow BES for an example channel.  Time range for inset in `c' is from 3880 to 3900~msec.  (e) Neutral beam power vs. time showing neutral beam turn off at 3999.42 msec.  Note that the slow BES signal drops and 360~Hz power supply oscillation in the fast BES signal disappears after the neutral beam is turned off.}
    \label{fig:sample_data}
\end{figure}

\begin{figure}
    \centering
    \includegraphics[scale=0.87]{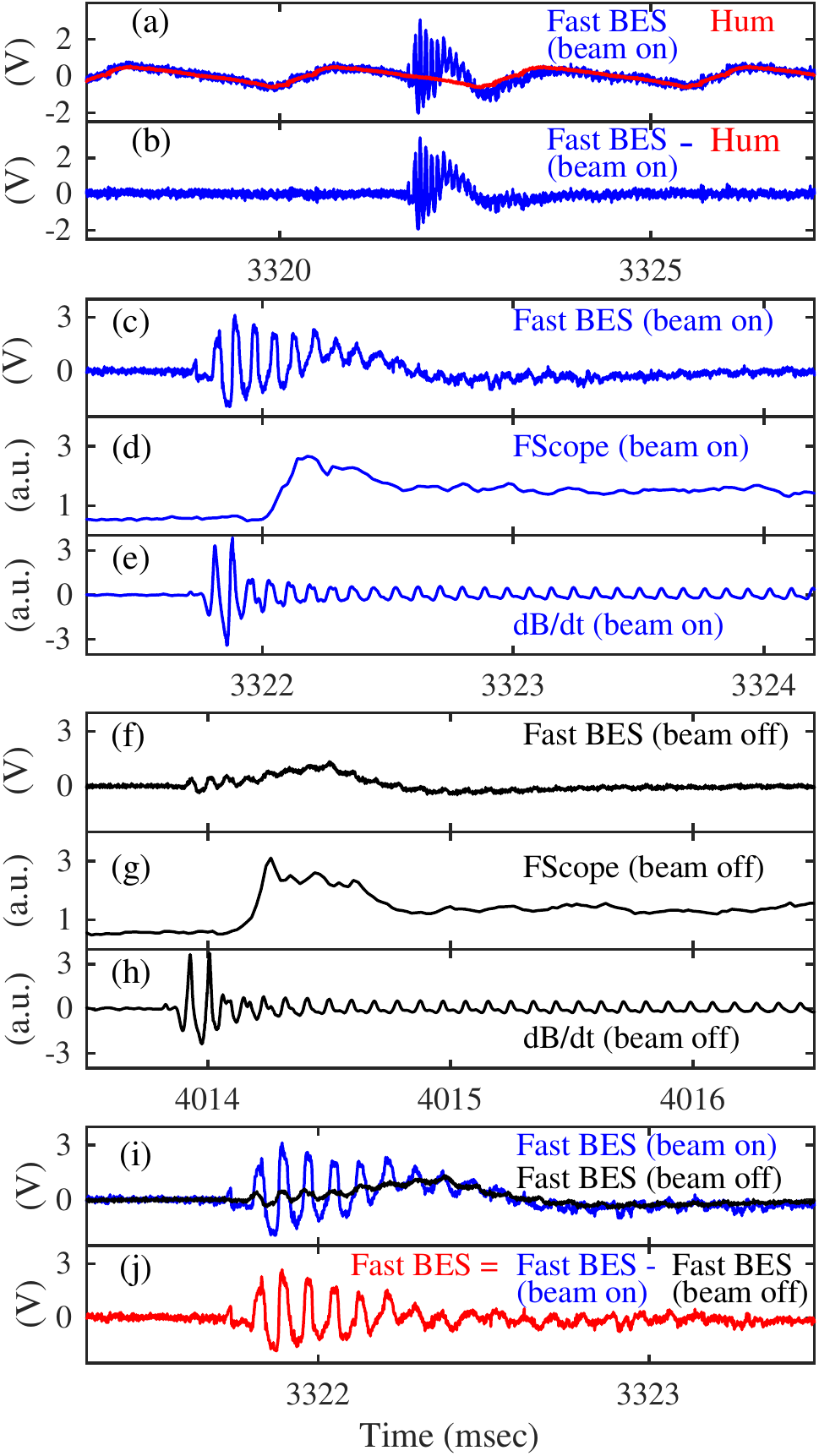}
    \caption{Demonstration of the processing routine for raw fast BES signal using channel 28 data as an example. Blue and black lines are used to represent data with neutral beam ``on" and ``off", respectively. The red is used for curves obtained after some mathematical operation. (a) Raw BES fast channel data where the power-supply hum in red is removed in (b). (c) Fast BES, (d) Filterscope, and (e) B-dot probe data corresponding to a sawtooth event when the neutral beam was on. (f) Fast BES, (g) Filterscope, and (h) B-dot probe data corresponding to a sawtooth event after the neutral beam is turned off at 3999.42 msec.  (i) Sawtooth signals before and after neutral beam turn-off aligned for edge light removal. (j) Sawtooth signal with edge light removed.}
    \label{fig:analysis_steps}
\end{figure}

Data from DIII-D shot number 176392 is shown in   Figure~\ref{fig:sample_data}, which shows a sequence of 5 sawtooth crashes during the plasma current flat-top, and which we will use to illlustrate the analysis chain. $T_{e,\rm{core}}$  exhibits a sawtooth pattern, rapidly falling from $\sim 3.2$ to $\sim 1.9$~keV at each crash, as seen in Figure~\ref{fig:sample_data}(a).  At each sawtooth crash, the B-dot probe (Figure~\ref{fig:sample_data}b) 
and fast BES  (Figure~\ref{fig:sample_data}c) also
show a burst of activity.  The slow BES signal containing background density information and the time variation of the power of the neutral beam used by BES diagnostic are shown in Figures~\ref{fig:sample_data}(d) and (e), respectively.

In the experiment, the neutral beam was switched off at 3999.42~msec. This leads to a rapid, but not complete, drop in the BES signal.   A crucial point is that BES continues to observe a response during sawtooth events even when the BES probe beam is off.  Additionally the fast BES signal has a $\approx$~360~Hz frequency component in the presence of the neutral beam current, which is due to power-supply ripple on the neutral beam, and which is absent
once the beam is off.  These features of the BES signal with and without the probe beam provide essential components for developing the data analysis routine.

We now describe how we process the BES data to obtain the localized density evolution during a sawooth event. (Figure~\ref{fig:analysis_steps}) 
The first processing step removes the $\approx 360~\rm{Hz}$ ``hum" caused by the neutral beam power supply. A simple filter cannot be used to remove the hum because the frequency of the light collected from the plasma has frequencies near 360~Hz. The hum is instead isolated with a digital comb filter obtained by averaging the signal with several shifted versions. Isolated ``hum" is shown by the red curve in Figure~\ref{fig:analysis_steps}(a), which is subtracted off in  Figure~\ref{fig:analysis_steps}(b).

Next, a comparison of the fast signal before and after the neutral beam turn-off in Figures~\ref{fig:analysis_steps}(c) and (f) shows that the fast BES observes a significant burst during a sawtooth crash even in the absence of the neutral beam.  While the BES information  is ordinarily localized by the probe beam, ultimately the field of view also goes through edge plasma regions which may also have large $D_\alpha$ emission.   Ideally, the edge emission is expected to be spectrally filtered as it is not doppler-shifted to the neutral-beam velocity, but the filtering is evidently not complete.  Second, it appears these sawteeth events are violent enough that they cause an extra burst of plasma and light at the edge, which is synced to the sawteeth events and slightly delayed by a few 100 microseconds..

We compared the fast BES data with the filterscope data in Figures~\ref{fig:analysis_steps} (d) and (g) to understand this.  The  filterscope does not sample the identical plasma volume viewed by the BES but does acquire light from the plasma edge.  Both the fast BES and filterscope exhibit a burst of signal following each sawtooth.  The bursts arise contemporaneously or shortly after a sawtooth crash, overlaps the measured core instability signal in time, and has roughly similar measured amplitude.  This suggests that a rapid expulsion of particles, caused by a sawtooth, creates a burst of edge light emission as they recycle from the vessel walls.

We adopted the following procedure to remove the edge light from the sawtooth events. The steps followed to remove the edge light from the fast BES are shown in Figures~\ref{fig:analysis_steps}(e), (h), (i) and (j).  
First, the sawteeth events exhibit a few characteristic patterns on the B-dot probes, likely related to toroidal differences from sawtooth to sawtooth.  Therefore, we first identify beam-on and beam-off sawtooth events which have a similar B-dot pattern.

Next, the beam-on and beam-off B-dot probe signals  (Fig.~\ref{fig:analysis_steps}(e) and (h)) are cross-correlated to find the lag time.  This lag time is used to align the fast BES signals, as shown in Figure~\ref{fig:analysis_steps}(i). We can observe that the initial fast oscillations early in the event are not matched between beam-on and beam-off cases, but that the $\sim$ms-time-scale ``pulse'' toward the end of the event is nearly identical in the two cases. This suggests that the initial fast oscillations in the beam-on event represent \textit{bona fide} core density oscillations, but the longer pulse is simply the edge light. Finally, the beam-off fast BES signal is subtracted from the beam-on fast BES to obtain the fast BES signal from core plasma shown in Figure~\ref{fig:analysis_steps}(j).    Henceforth, this fast BES signal from the core is referred to as $V_{\rm{f}}$. Note that $V_{\rm{f}}$ is not symmetric about zero but exhibits a positive skewness, so a simple  bandpass filter is not appropriate to remove the edge light.

The slow BES signal is shown in Figure~\ref{fig:sample_data}(d).  The signal drops after the beam turn-off, but the signal does not go to zero. We checked the entire time series of the slow BES signal and observed that the signal goes to zero only in the absence of plasma. This indicates that in the presence of plasma, the BES photodiodes collect some light  primarily from the residual edge recycling and visible bremsstrahlung, even in the absence of the neutral beam. The slow BES is recording that signal. The average light from the sawtooth inversion region due to neutral beam fluorescence is isolated by subtracting the slow BES signal averaged over a sawtooth period after beam turn-off from the signal when the beam is on. This average signal from the slow BES channel is referred to as $V_{\rm{s}}$ and is used to calculate light fluctuation due to sawteeth from the inversion region in the following subsection.

\subsection{Conversion of raw BES signals to $\delta I/I$}

The relative core light emission variations,  $\delta I / I_0$, is obtained from $V_{\rm{f}}$ and $V_{\rm{s}}$ using the expression,
\begin{equation}
    \frac{\delta I}{I_{0}}=K_{\rm{s}} \frac{V_{\rm{f}}}{V_{\rm{s}}},\label{Eq:deltaI_byI_calc}
\end{equation}
where $K_{\rm{s}}$ is a standard calibration factor, which depends on the channel and was
obtained by checking the relative gain of the fast and slow channels on a test bench.

\begin{figure}
    \centering
    \includegraphics[scale=0.85]{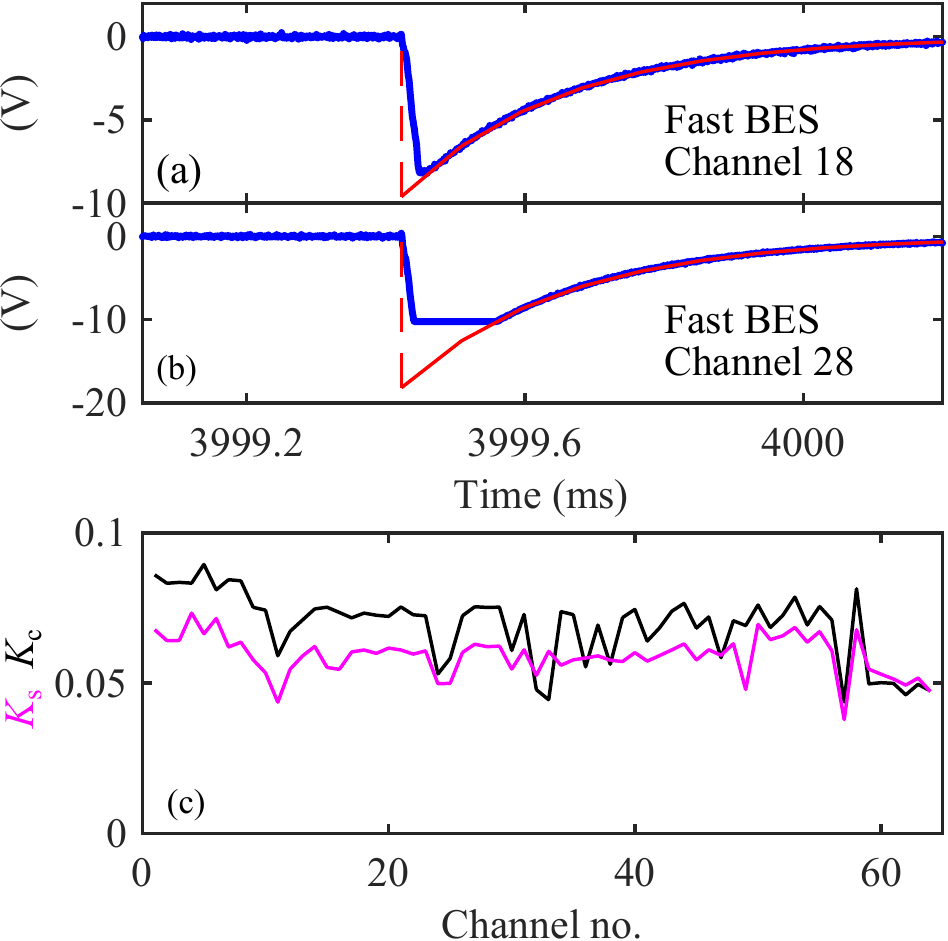}
    \caption{Fast BES signals at neutral beam turn-off.  Blue lines show the time variation of fast BES signal at beam turn-off in channels (a) 18 and (b) 28. An exponential fit to fast BES data during the recovery phase after beam turn-off is shown by the continuous red curves. The red dashed curve represent the change in fast BES at beam turn-off in the absence of saturation and finite response time of the electronics. (c) Calibration factor supplied by BES group and calculated using Equation~\ref{Eq:calib_fac_det} are given by magenta and black curves, respectively.  }
    \label{fig:calib_factor}
\end{figure}

We have verified the calibration factor for calculating $\delta I/I_{0}$ using the neutral beam turn-off events for  an \textit{in situ} calibration of the BES system. The calibration also complements the standard BES calibration, $K_{\rm{s}}$, as it provides  a direct time-domain calibration for  the channels in the lower frequency regime relevant for sawteeth events.

 At the beam turn-off event, the light onto the 
BES photodiodes undergoes an abrupt downward
step, as the core light (neutral beam fluorescence)
is removed, leaving only the edge light 
after turn-off.   The
slow arm observes this as an
downward step function of magnitude $V_s$ 
as defined and discussed in the
previous section.
The fast channel observes the same step function
through its $RC$ high-pass filter circuit,
which leads to a downward step followed by
an exponentially-decaying recovery to $0$.
We fit the response to the functional form,
\begin{equation}
    V_{\rm{f}}(t > t_{t-o}) = -\Delta V_{\rm{f},t-o} \exp(- (t - t_{t-o}) / \tau) \label{Eq:RC_response}
\end{equation}
where $t_{t-o}$ is the beam turn-off time and
$\tau$ is the high-pass filter time constant.
Since, in principle, the identical current
step is applied to both fast and slow channels,
the voltage steps observed on the fast and
slow channels and the calibration coefficient
are related by the equation,
\begin{equation}
    1=K_{\rm{c}}\frac{\Delta V_{\rm{f}, t-o}}{V_{\rm{s}}},\label{Eq:calib_fac_det}
\end{equation}
where $K_{\rm{c}}$ is a new calibration coefficient defined by this relation.

In Figures~\ref{fig:calib_factor} (a) and (b), blue curves show a zoomed-in view of fast BES data for channels 18 and 28, respectively. 
We note that the Fast BES signals are observed to take a finite amount of time to decrease to the lowest value at beam turn-off, due to finite frequency response.  Also, some channels were found to saturate, in which case we fit to the non-saturated portion of the data.

The calculated values of $K_{\rm{c}}$ for various channels are compared with $K_{\rm{s}}$ in Figure~\ref{fig:calib_factor}(c), and both calibration factors are found to agree within $\sim20\%$.

\begin{figure}
    \centering
    \includegraphics[scale=0.85]{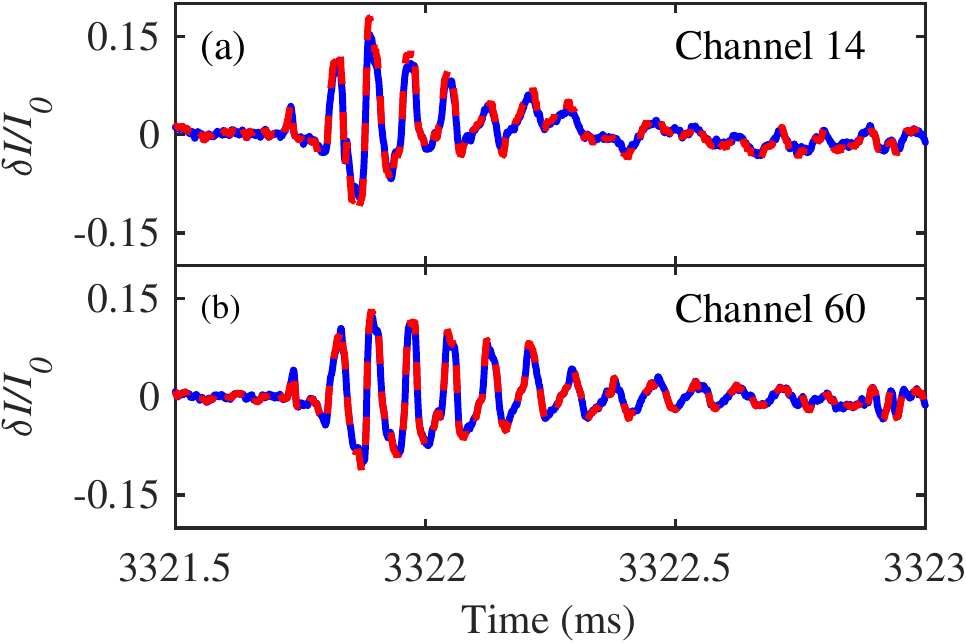}
    \caption{Time variation of $\delta I/I_{0}$ measured by channel (a) 14 and (b) 60. The continuous blue and broken red lines were obtained using $K_{s}$ and $K_{c}$, respectively.  }
    \label{fig:delta_I_by_I_time}
\end{figure}

Figure~\ref{fig:delta_I_by_I_time} shows $\delta I/I_{0}$ calculated using $K_s$ and $K_c$. The wave-trains obtained using $K_{\rm{s}}$ and $K_{\rm{c}}$ are similar. Henceforth, we have used $K_{s}$ to determine $\delta I/I_{0}$ from $V_{\rm{s}}$ and $V_{\rm{f}}$.

 Finally, we note that the discussion is also useful for a quick cross-check of the data ``by eye.'' Adopting a ``$K_c$'' calibration, i.e., adopting a calibration in Eq.~\ref{Eq:deltaI_byI_calc} using $K_c$, and  inserting Eq.~\ref{Eq:calib_fac_det}, one obtains,
 \begin{equation}
       \frac{\delta I}{I_{0}} = 
       \frac{V_{\rm{f}}}{\Delta V_{\rm{f, t-o}}},
       \label{Eq:deltaI_byI_new}
 \end{equation}
 which is useful because it does not involve the  slow channel.   Secondly, one can then immediately read off from raw data, for example  Figure~\ref{fig:sample_data}c, that the $\delta I/I_{0}$ during sawtooth crash events are $\sim10-20$\% of the downward step during the turn-off event (for channels that do not saturate), which directly supports the inferred $\delta I/I_0$ from the full calibration (Fig.~\ref{fig:delta_I_by_I_time}).

\subsection{Conversion of $\delta {I}/{I}_{0}$ to density variation}

The conversion of the BES intensity variations $\delta I / I_{0}$ to plasma density variations is the final step, and per standard procedure, requires an atomic physics model\cite{FonckRSI1990,HutchinsonPPCF2002}.   The intensity of light emission  depends on the number of collisionally excited neutrals in the neutral beam undergoing $\rm{n}=3$ to $\rm{n}=2$ transitions.  The fractional population of $\rm{n}=3$ state depends on the local plasma density,  temperature, $T$, impurity ions, $Z_{\rm{eff}}$, and neutral beam energy, $E_{\rm{beam}}$\cite{FonckRSI1990}. The relationship between $\delta I/I_{0}$ and $\delta n_{e}/n_{e0}$ taking into consideration the intricacies of the mechanism of emission from the neutral beam is given by
\begin{equation}
    \frac{\delta n_{e}}{n_{e0}}=C (n_{e}, T, Z_{\rm{eff}}, E_{\rm{beam}}) \frac{\delta I}{I_{0}},\label{Eq:deltan_by_n}
\end{equation}
where $C$ is the proportionality factor\cite{FonckRSI1990}.  This linearized version is valid for small density fluctuations where $C$ is replaced by a constant number\citep{HutchinsonPPCF2002,FonckRSI1990}.

\begin{figure}
    \centering
    \includegraphics[scale=0.85]{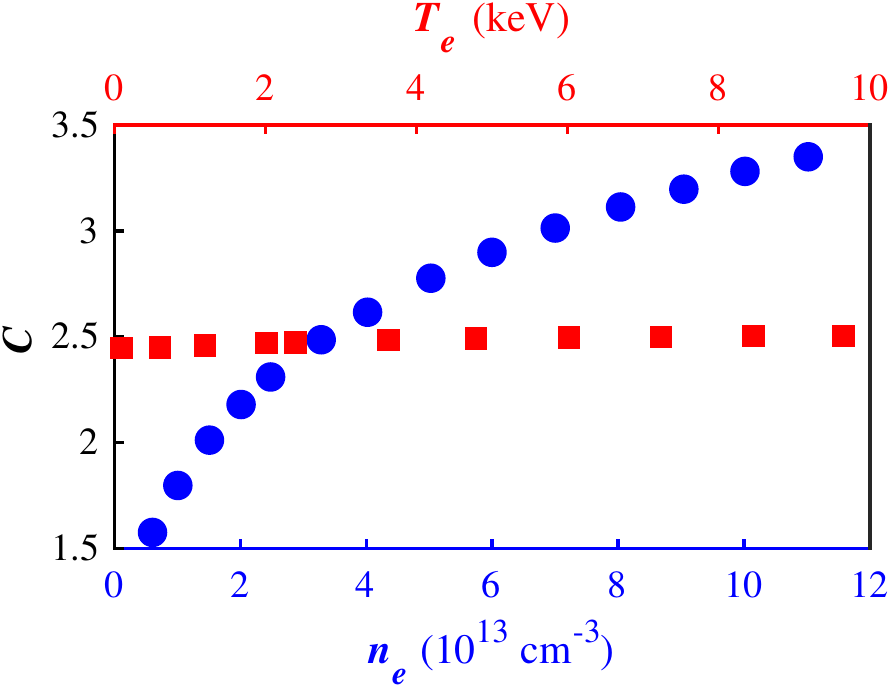}
    \caption{Variation of the proportionally factor $C$ with density and temperature. Density and temperature were held constant at $n_{e}=3.3\times10^{13}~\rm{cm^{-3}}$, and $T_{e}=1.7~\rm{keV}$, for determining the $C$ vs.\ $T_{e}$, and $C$ vs.\ $n_{e}$ curve, respectively.  }
    \label{fig:C_dependence}
\end{figure}

We calculated the dependence of $C$ on plasma parameters, shown in Fig.~\ref{fig:C_dependence}. The red squares in Figure~\ref{fig:C_dependence} show that $C$ does not vary with $T_{e}$ in our operation regime.  However, $C$ does depend on the $n_{e}$, which indicates that for large $n_{e}$ fluctuations the non-linearity should be preserved.
The effect of $E_{\rm{beam}}$ and $Z_{\rm{eff}}$ on $C$ are ignorable\cite{FonckRSI1990}. Therefore, for our regime of operation, Equation \ref{Eq:deltan_by_n} can be written as 
\begin{equation}
    \frac{\delta n_{e}}{n_{e0}}\approx C(n_{e}) \frac{\delta I}{I_{0}}.
\end{equation}

\begin{figure}
    \centering
    \includegraphics[scale=0.85]{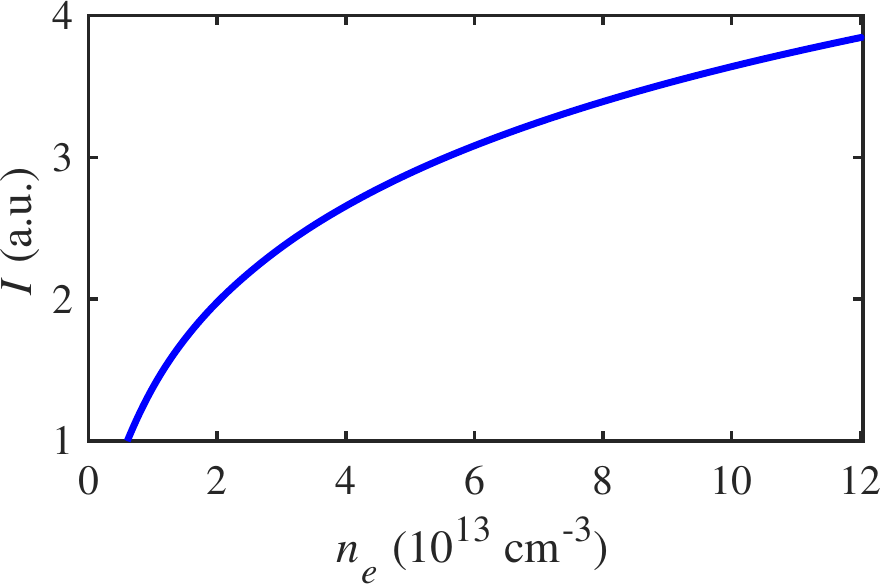}
    \caption{Dependence of intensity on density}
    \label{fig:I_vs_n}
\end{figure}

In order to handle density variations beyond the linear regime, we obtain
the full non-linear $n_{e}$ vs.\ $I$ relationship (starting from tabulated
data of Eq.~\ref{Eq:deltan_by_n} shown in Fig.~\ref{fig:C_dependence}) by separating variables and integrating, 
\begin{equation}
    \int^n \frac{dn'}{C(n') n'} = \int^I \frac{dI'}{I'}.
\end{equation}
The resulting calibration curve $I(n_{e})$ for the present operating
parameters is plotted in Figure~\ref{fig:I_vs_n}.
We verified that the resulting curve could be re-linearized to
reproduce Figure~\ref{fig:C_dependence}.

Figure~\ref{fig:delta_n_by_n}(i) shows $\delta n_{e}/n_{e0}$ obtained from $\delta I /I_{0}$ in Figure~\ref{fig:delta_I_by_I_time}(a) using the $I$ vs.\ $n_{e}$ calibration curve. To use the curve, a central operating point $(n_{e0}, I_0)$ must be chosen that
corresponds to the point when $\delta I = 0$.  
Here, we use $n_{e0} = 3.5\times 10^{13}~\rm{cm^{-3}}$, which was obtained from a time-average 
of a Thomson scattering density channel close to the $q=1$ surface.
Then $I(t)$ was obtained using first, 
\begin{equation}
I(t)=I_{0}\times \bigg\{1 + \frac{\delta I}{I_{0}}(t) \bigg\},
\end{equation}
after that, $I(t)$ was converted to $n_{e}(t)$ using the $I$ vs.\ $n_{e}$ calibration curve. Finally, the time variation of $\delta n_{e}/n_{e0}$ was calculated using,
\begin{equation}
\frac{\delta n_{e}(t)} {n_{e0}}=\frac{n_{e}(t) - n_{e0} } {n_{e0}}.    
\end{equation}
We used the  same $n_{e0}$ for all BES channels. Furthermore, $\delta n_{e}/n_{e0}$ data from channels 5, 19, 21, 26, 29, 36 and 59 were found to have some systematic issues due to the hardware. One channel is dead and the transducer in other channels might have a nonlinear response to neutral beam fluorescence.   We replaced the data of those channels by the time series obtained by averaging data from neighbouring channels.

\subsection{Density variation during sawtooth events}

\begin{figure}
    \centering
    \includegraphics[scale=0.25]{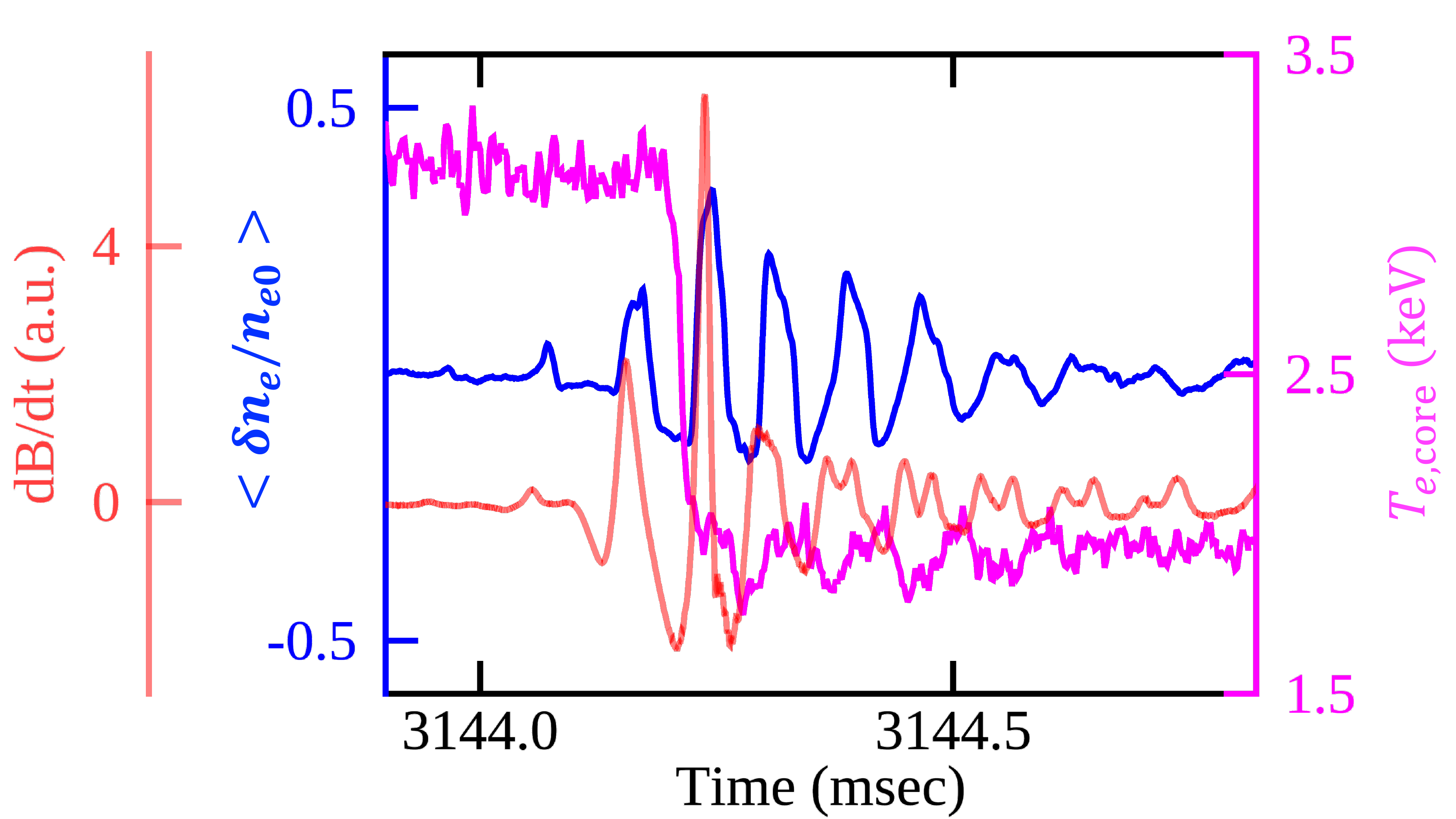}
    \caption{Time variation of BES array-averaged density, $\langle \delta n_{e} (t) /n_{e0} \rangle$, $dB/dt$, and $T_{e,\rm{core}}$ in blue, light red, and magenta, respectively. }
    \label{fig:avg_density}
\end{figure}

Using the techniques above we now present  the density evolution measured by the BES during a sawtooth event. We first note that the large oscillation discussed above and shown in Figure~\ref{fig:delta_I_by_I_time} is well-correlated across the entire BES array. Figure~\ref{fig:avg_density} shows the time evolution of the array-average $\langle \delta n_{e} (t)/n_{e0} \rangle$, where we have taken the average over all 64 BES channels. (We note that the BES array, deployed here over an area of 20 $\times$~8~cm, is still a small fraction of the entire plasma cross section, 
i.e. see Fig.~\ref{fig:setup}.)
The magenta curve shows the associated time variation of $T_{e,\rm{core}}$, as measured by a core ECE channel.  The density exhibits a large amplitude oscillation at a frequency of  $\sim 13~\rm{kHz}$ that starts just before the temperature crash  and persists for a few cycles afterwards. This density variation coincides with large magnetic fluctuations (Figure~\ref{fig:avg_density}).  The good correlation of this structure across the whole array indicates this structure is at an equal or larger scale to the array, i.e. a large-scale mode in the plasma. This oscillation is consistent with a helical $(1,1)$ mode growing during a sawtooth  event, where the oscillation is  due to the rotation of the mode (with the plasma) past the fixed measurement location. The density oscillations are relatively large, with magnitude up to $\langle \delta n_{e}/n_{e0} \rangle \sim 0.2-0.4$ indicating a significant modification of the  plasma density profile during the event. The large density oscillations are observed to  persist for $\sim$5 plasma rotations, after which they progressively decay.

\begin{figure*}
\includegraphics[scale=1.05]{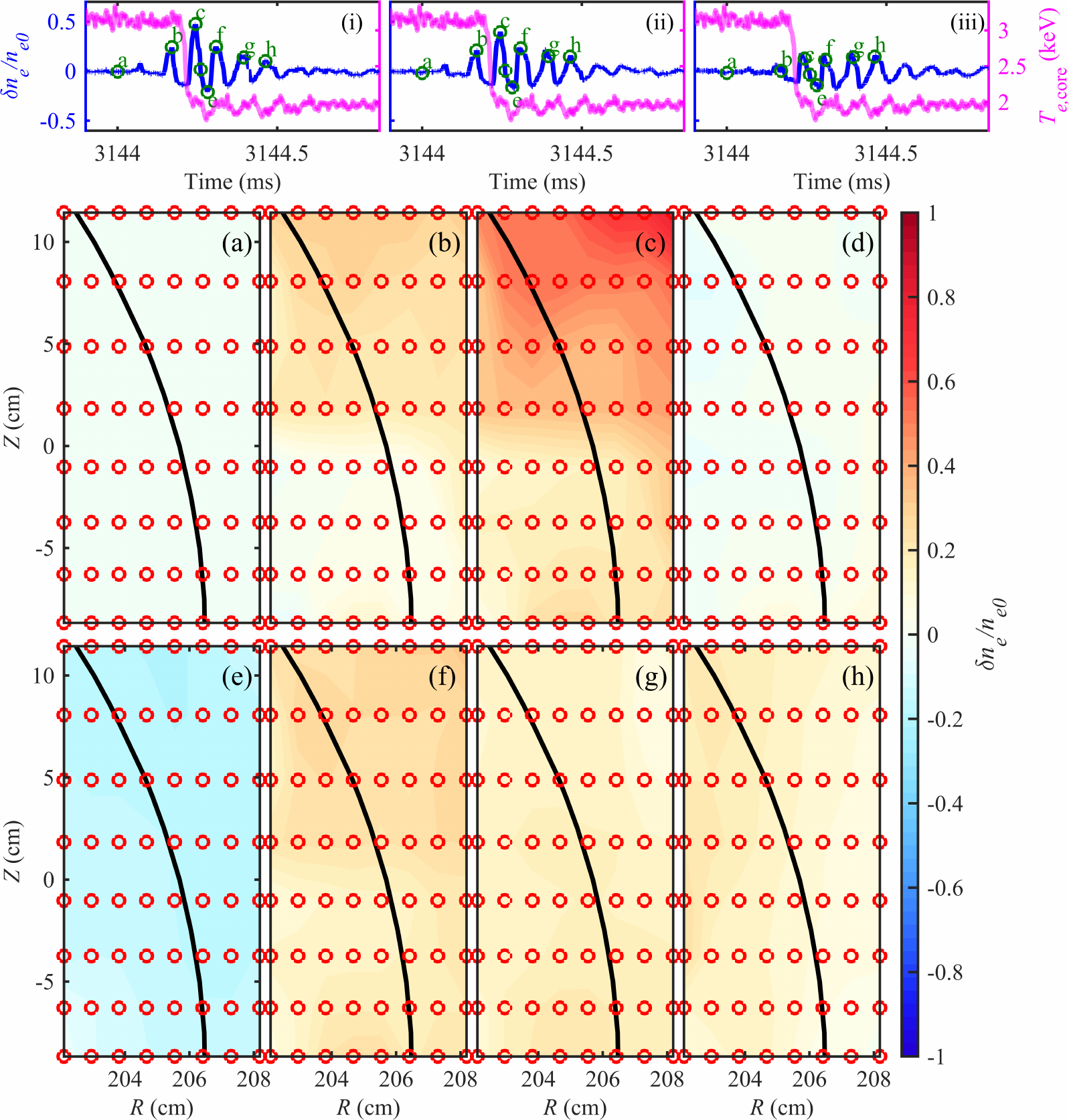}
\caption{\label{fig:delta_n_by_n} 
2-D BES observations during a sawtooth event. Magenta coloured line in i, ii, and iii show the temporal variation of core electron temperature. Blue lines show the time variation of $\delta n_{e}/n_{e0}$ measured by (i) channel 14 located to the right of the inversion layer, (ii) channel 20 at the inversion layer, and (iii) channel 35 located to the left of the inversion layer, where $n_{e0}=3.5\times 10^{13}~\rm{cm^{-3}}$. The circles in green in the time series data are the time instants for which we have plotted 2-D images of $\delta n_{e}/n_{e0}$ in the $RZ$ plane near the $q=1$ surface. Labels of the 2-D images specifies those time instants. For example, figure (a) shows the spatial profile of $\delta n_{e}/n_{e0}$ before the crash and this time instant is marked as `a' in figures i, ii and iii. The black line on the 2-D images represent the contour of $q=1$ surface. The red circles show the location of the center of the BES channels. Note that before a sawtooth crash,
$\delta n_{e}/n_{e0}$ is nearly uniformly zero across 
the $R$-$Z$ plane. 
However, during a crash $\delta n_{e}/n_{e0}$ develops a significant spatial variation across the $q=1$ surface.}
\end{figure*}

Local variation of density during a sawtooth event is shown in Figure~\ref{fig:delta_n_by_n} using  1-D and 2-D plots. The blue curves in Figures~\ref{fig:delta_n_by_n}(i), (ii), and (iii) show the time variation of the density for three representative channels, (i) outboard of the sawtooth inversion layer, (ii) at the inversion region, and (iii) inboard of the  inversion layer, respectively. The amplitude of the density oscillation is higher on the outboard side than on inboard of the inversion layer.

The 2-D color plots in Figure~\ref{fig:delta_n_by_n} show the spatial profiles of the density at different instants of  time. The 2-D data was passed through a median filter before making those color plots. The black curve on the 2-D color plots shows the contour of the $q=1$ surface, and the color represents the magnitude of $\delta n_{e}/n_{e0}$, where $n_{e0}=3.5\times10^{13}~\rm{cm^{-3}}$. For example, Figure~\ref{fig:delta_n_by_n}(a) shows that density is uniform in the sampled part of the $R$-$Z$ plane much before the crash in $T_{e,\rm{core}}$.

The temporal variation of the spatial profile in density during a sawtooth oscillation is studied by making a video of the time evolution of the 2-D  density data.  Figures~\ref{fig:delta_n_by_n}(b) - (h) show snapshots of the 2-D variation of density at the times indicated in (i)-(iii), which correspond to times of a few maxima, minima, and zero-crossings in $\delta n_{e}/n_{e0}$.

Interestingly, in addition to the array-averaged density variations described above, significant in-plane density non-uniformities are also observed. Figure~\ref{fig:delta_n_by_n}(b) shows that $\delta n_{e}/n_{e0}$ is non-uniform in the $R$-$Z$ plane at the onset of the crash in $T_{e,\rm{core}}$.  
The density increases in the upper right region of the $R$-$Z$ plane and decreases in the lower left region.  The density varies in the plane from $\sim 3.7\times 10^{13}$  to $4.6\times 10^{13}~\rm{cm^{-3}}$. As the crash in $T_{e,\rm{core}}$ continues, the density non-uniformity increases significantly. 
The  density in the $R$-$Z$ plane ranges from  $\sim4\times10^{13}$ to $\sim6\times10^{13}~\rm{cm^{-3}}$ at the latter end of the crash as seen in Figure~\ref{fig:delta_n_by_n}(c). The latter oscillations show much more uniform $\delta n_{e}/n_{e0}$ (panels g and h), as do the times of the minima of $\delta n_{e}/n_{e0}$ (panel e). We provide a caveat that the fine-structure on the array may still be  influenced by the channel-to-channel variations in response, and therefore we do not discuss the fine-scale structures here.  Understanding the fine-scale 2-D  structures will be in the scope of future work.

\section{\label{sec:discussion}Discussion and Summary}

We have presented a technique allowing the localized fast measurement of density in the $R$-$Z$ plane near the $q=1$ sawtooth inversion region during a sawtooth crash. We have developed a comprehensive method for analyzing the BES data considering various pitfalls.  Since the measurement of the  density is sensitive to the calibration of the BES diagnostic, a novel technique was developed to cross-verify the channel-to-channel calibration during a plasma shot. Our technique does not require any additional measurement on the test bench.

Intense $\rm{D}_{\alpha}$ emission due to edge recycling caused by sawtooth oscillation makes the use of standard spectroscopic filtering techniques insufficient for removal of edge light. Traditional data analysis technique like Fast Fourier Transform (FFT) cannot be used because
there is not a significant frequency separation between core and edge light signals.  Therefore, we developed a method for isolating and removing undesired edge light from the BES data by comparing beam-on and beam-off events.    

In experiments where light fluctuations due to neutral beam fluorescence are small, $\delta I/I_{0}$ is related to $\delta n_{e}/n_{e0}$ by a constant of proportionality. However, light fluctuations are large during a sawtooth crash, and the neutral beam emission depends nonlinearly on density.  Therefore, we inverted the light intensity to density considering the complete nonlinear dependence.

A total of 64 BES channels in $8\times8$ configuration spanning a $8~\rm{cm}$ (radial) $\times ~20~\rm{cm}$ (poloidal) area across the sawtooth inversion layer was used to make density  measurements. A large amplitude (1,1) mode is observed in the density data. The mode starts almost at the onset of the crash.   The maximum amplitude of this mode  is, $\langle \delta n_{e}(t)/n_{e0}\rangle_{\rm{max}}\approx$ 0.4 (Figure~\ref{fig:avg_density}). This mode persists for a few cycles even after the crash.

Multiple localized density measurements in the $R$-$Z$ plane show that at the onset of the sawtooth crash, the density becomes spatially inhomogeneous near the sawtooth inversion layer with $\Delta n_{e}(R,Z)/n_{e}\sim 0.2$. This spatial nonuniformity in density significantly increases at the latter end of the crash; $\Delta n_{e}(R,Z)/n_{e}~\sim 0.4$. 

In addition to the measurements reported in this article, we have done other experiments also. In one of those experiments, the BES diagnostic was moved to sample the plasma further to the right and to the left of the $q=1$ surface. We observed $\delta n_{e}/n_{e0}$ to be strong near the $q=1$ surface and decrease further away on either side. We will report these results elsewhere.

The spatial variation of density reported in this article may be related to  guide field reconnection or may indicate some other MHD phenomena.  We are making comparisons of the measured density variation with predictions of various models, which will be reported in future publications.

\begin{acknowledgments}
This material is based upon work supported by U.S. Department of Energy, Office of Fusion Energy and Sciences through DIII-D Frontier Science program using DIII-D National Fusion Facility, Max Planck Princeton Center for Plasma Physics, and partially funded under awards DE-FC02-04ER54698, DE-FG02-08ER54999, DE-AC0204CH11466, and  DE-AC02-09CH11466. The United States Government retains a non-exclusive, paid-up, irrevocable, world-wide license to publish or reproduce the published form of this manuscript, or allow others to do so, for United States Government purposes. 
\end{acknowledgments}

\vspace{20px}


\bibliography{apssamp}

\end{document}